\documentstyle[aps]{revtex}

\begin{document}
\draft
\title{Visible light emitting devices with Schottky contacts on 
silicon nanocrystals}
\author{S. Fujita and N. Sugiyama}
\address{Toshiba Corporation, Research and Development center, 
Advanced Research Laboratory 1, Komukai-toshibacho, Saiwaiku, 
Kawasaki, Japan, 210-8582}
\date{\today}
\maketitle
\begin{abstract}
     We have fabricated light emitting diodes (LEDs)
 with Schottky contacts on Si-nanocrystals formed by
 simple techniques as used for standard Si devices. 
Orange electroluminescence (EL) from these LEDs could be
 seen with the naked eye at room temperature 
when a reverse bias voltage was applied. 
The EL spectrum has a major peak with a photon 
energy of 1.9 eV and a minor peak with a photon 
energy of 2.2 eV. Since the electrons and holes are 
injected into the  radiative recombination centers 
related to nanocrystals through avalanche breakdown,
 the voltage needed for a visible light emission is 
reduced to 4.0 - 4.5 V, which is low enough to be 
applied by a standard Si transistor. 
\end{abstract}
 
\medskip

\narrowtext
   There have been expectations that high-efficiency
 light emitting diodes (LEDs) using silicon-(Si-) 
related materials can be realized for monolithic 
optoelectronic integrated circuits (ICs). 
Various types of LED based on Si-related materials 
have been tried to date. Of Si-related materials 
investigated, porous Si (PS) has been studied most 
actively. The quantum efficiency and stability of 
LEDs based on PS have been gradually improved.\cite{ref1}
 Recently, Si nanocrystals (Si-NC) have also been 
studied for their light-emitting characteristics. 
Si-NCs can be formed using methods which fit in better 
with standard silicon technology as compared with PS, 
since electrochemical processes are not needed. 
Based on the previous studies, 
it is predicted that the optical band gap is 
increased\cite{ref2} and the probability of 
radiative recombination for excitons is 
enhanced\cite{ref3} with a decrease in the size 
of nanocrystals owing to the quantum confinement effect. 
Significant increases in oscillator strength are 
also predicted at the very small dimensions 
of NCs.\cite{ref4} Visible photo-luminescence 
(PL) with high efficiency has been reported with 
Si-NCs formed by various methods: chemical 
vapor deposition using silane or disilane gas 
resolved by microwave plasma,\cite{ref5} 
laser\cite{ref6}, or thermal reactions;\cite{ref7} 
ion implantation of Si into silicon oxide 
films followed by post-annealing;\cite{ref8} 
and crystallization of amorphous Si (a-Si) 
films.\cite{ref9} The reported photon energy of 
the PL peak spectra is 1.5 to 1.9 eV (red) for 
most Si-NCs. 
Blue PL emission with a peak photon energy of 3.2 eV 
has also been obtained from Si-NCs formed by 
crystallization of amorphous Si films.\cite{ref9} 
The luminescence related to defects in SiO2 with a 
peak photon energy more than 2 eV has also been reported.\cite{ref8} 
   The number of the reports on electroluminescence
 (EL) from the LEDs\cite{ref10}\cite{ref11}\cite{ref12} 
is very small as compared with that 
for PL from Si-NCs. This is principally 
because the carrier injection induced 
electrically is more difficult than that induced 
optically. In the investigations  reported, 
visible EL was obtained from Si-NCs in SiO2 matrix formed by
 chemical vapor deposition (CVD) \cite{ref10} or by 
Si+-implantation into SiO2;\cite{ref11} and from a 
multi-quantum-well (MQW) composed of silicon-nitride(SiNx)/Si-NC
 formed by plasma enhanced CVD and laser annealing.\cite{ref12} 
The EL in these cases was strong enough to be seen with the naked eye. 
These results indicate that there is promise in using Si-NCs to realize 
LEDs based on Si-related materials. However, 
as yet the operating voltage is still rather high (10~25 V or more)
 for the devices to be driven by standard Si-ICs. This is because 
the Si-NCs in these cases are sandwiched between insulating layers 
or embedded by insulating layers, through which the carriers must 
pass by direct tunneling. This high voltage must be reduced in 
order to use LEDs based on Si-NCs practically. A simple structure 
which can be fabricated easily by standard Si-IC technology is 
also desired for the LED.
   In this study, we have fabricated simple LEDs with Schottky contacts
 on Si-NCs and without the insulating layers. Visible EL has been observ
ed
 form this LED. The EL is strong enough to be seen with the naked eyes
 at room temperature and with a relatively low operating voltage.
   The Si-NCs were formed as follows by the crystallization of a-Si . 
The p-type Si wafer had a diameter of 150 mm and low 
resistivity (0.01 ohm cm). The native oxide was removed 
by etching before the deposition of a-Si. A molecular beam of Si, 
formed by resolving disilane using a tungsten cracking 
heater,\cite{ref13} was deposited onto the Si substrate 
at room temperature.
 The a-Si was converted into nanocrystals by rapid heating to 
700 degree C for one or three minutes in pure oxygen gas. 
During this heating, oxidation of silicon occurs simultaneously. 
The oxidation is effective to stabilize the surface. 
It was known in advance that the thickness of oxide formed 
on a crystalline Si substrate for one and three minutes of 
oxidation was about 0.5 and 0.7 nm, respectively, in the oxidation 
furnace used. PL measurements for these samples were performed 
using the Ar+ laser (458nm) at room temperature. 
The structure of these samples was observed using a high-resolution
 transmission electron microscope (HR-TEM). Metal contacts 
composed of titanium and gold (Ti/Au) were formed by 
electron beam (EB) evaporation on the back of the substrate. 
Ti/Au contacts 100 x 100 mm2 in area were formed on the surface 
of the Si-NCs by the lift-off process. Device isolation was not 
employed. Figure 1 shows a cross-sectional scheme of the resulting 
device. For reference, p-type Si substrates were also annealed 
in oxygen gas and the devices with Schottky contacts were 
fabricated on them. The current-voltage (I-V) characteristic 
was measured at room temperature. EL was measured at room temperature 
using a monochrometer and a photomultiplier. 
   At room temperature, the PL signal with peak photon energy of 
1.9eV was obtained from the a-Si layers that experienced one minute 
of oxidation. On the other hand, there were no detectable PL signals 
from either the a-Si layers without annealing or the samples 
that experienced three minutes of oxidation. Figure 2 shows a 
cross-sectional HR-TEM lattice image of an a-Si layer that 
experienced one minute of oxidation. Tiny Si-nanocrystal islands
 of almost hemispherical shape and 2-3 nm wide are seen
 in the a-Si layer. The presence of stacking faults between 
the islands and the substrate indicates that the islands did not 
exist prior to deposition of the a-Si layer. It is possible that 
the other nanocrystals of dimensions less than about 2nm was formed 
in a-Si layer, since lattice image of such small crystals can hardly
 be obtained even by HR-TEM. The surface of the a-Si layer may be 
oxidized, though the oxide layer cannot be distinguished from the 
a-Si layer owing to the low contrast between the two in the TEM image. 
A previous report indicates that 2-3nm rectangular-shaped nanocrystals 
were formed by a similar method and that strong blue PL was 
observed.\cite{ref14}  In the case of a-Si layers that experienced 
three minute of oxidation, it was reveled that most of the a-Si was 
converted into large crystals of Si, whose surface was slightly oxidized
.
 Thus, it was founded that the origin of PL was related to the 
formation of Si-NCs.
   The I-V characteristics of these Si-NC devices show rectifying 
behavior. Figure 3 shows a typical I-V characteristic measured 
for a device that experienced one minute of oxidation. 
Devices which experienced three minutes of oxidation exhibited 
smaller forward and reverse currents. The I-V characteristic of 
each device was stable and changed little after repeated measurements.
   Orange EL emissions from Si-NC devices that experienced one minute 
of oxidation were clearly seen with the naked eye at room temperature 
when a reverse bias voltage was applied. The critical reverse current 
for the observation of light emission with the naked eye was 30 to 
40 mA at an applied voltage of 4.0 to 4.5V. In contrast, it was not 
possible to obtain visible EL emissions from either Si-NC devices that
 experienced three minute of oxidation or the reference devices 
without a-Si layers at room temperature. These results indicate 
that the visible EL is related to the existence of Si-NCs formed in a-Si
.
   Figure 4 shows a typical EL spectrum for a Si-NC device. 
This spectrum was measured at room temperature under a reverse 
bias of 5.0V and at a current of 60 mA. Two peaks are visible. 
The peak wavelength, equivalent photon energy and full-width at 
half-maximum (FWHM) are 650 nm, 1.9 eV and 110 nm for the major 
peak (peak A), and 570 nm, 2.2 eV and 30 nm for the minor peak 
(peak B), respectively.The FWHM of the major peak (peak A) is 
relatively small as compared with that of a PS EL spectrum.\cite{ref15} 
The peak wavelength and FWHM of the main peak is close to that 
obtained from a multi-quantum well (MQW) of (SiNx)/Si-NC.\cite{ref12} 
In the report of this MQW, it is thought that the orange EL is due to 
radiative recombination between the states quantized by the quantum 
confinement at the Si-NCs. For the tiny Si-nanocrystal islands seen 
in the TEM image of Fig.2, their structure may be not effective for 
the quantum confinement. However, in the case that nanocrystals which 
are invisible in this TEM image are formed in SiOx near the surface, 
the quantum confinement is possible. In contrast, it has been reported 
that the red PL (1.5-1.9eV) in the case of the oxidized Si-NC 
particles\cite{ref5}\cite{ref6}\cite{ref7} is due to the localization 
of excitons or carriers at the Si/SiOx interface.\cite{ref16} The peak 
energy for peak B (2.2eV) is close to that for PL spectrum of oxidized 
Si-NC particles reported in another article;\cite{ref17} 
where its origin seems to be Si/SiOx interface states; however, 
the details are unknown. In order to clarify the origin of these EL peak
s, 
it is necessary to examine the details of luminescence such as 
temperature dependence of both peak energy and peak intensity, 
the size dependence of peak energy, and the time-resolved 
characteristics of light emission. These will be studied 
in the near future. Although the origin of EL is unclear at this stage, 
it is evident that these light emissions are deeply related to 
the formation of the Si-NCs. It should be noted that the spectrum shown
 in Fig.4 is quite different from that produced by a reverse-biased 
p-n junction of silicon, where the spectrum spans the entire visible
 range and the emitted light is white.\cite{ref18}
   The observation that EL is emitted only under reverse bias is unique.
 Though similar characteristics have been obtained for a LED based
 on germanium-NCs embedded in SiO2,\cite{ref19} such behavior 
is quite different from that reported in the past for those on PS or Si-
NCs.
 EL was obtained when avalanche breakdown occurred, as shown in Fig.3, 
where the electrons and holes that were produced by impact ionization 
have enough energy to be injected into the emission centers at NCs. 
On the contrary, when under forward bias, the injection of carriers
 into these levels is impossible, since the energy of carriers outside
 the nanocrystals is low. Since avalanche breakdown easily occurs in our
 devices even with a low bias voltage, the operation voltage for 
visible EL can be reduced to 4.0 to 4.5 V, which is so low that 
these diodes can be driven by a standard Si bipolar transistor. 
It is surprising that the volume of the light-emitting layer for 
our devices is smaller by 30 to 500 times than those of the other 
cases where a bright EL is visible to the eye.\cite{ref11}\cite{ref12}  
It will also be possible to reduce the operating voltage of our 
devices further and to increase EL efficiency by optimizing the device 
structure. For this purpose, the use of avalanche breakdown in 
a p+/NCs/n+ structure will be also effective. Thus, the LED that 
we demonstrated is a promising device to realize a monolithic 
optoelectronic ICs. 
   In summary, we have fabricated LEDs with Schottky contacts 
on Si-NCs formed by a simple method. These LEDs give EL that can
 be seen with the naked eye at room temperature. The EL spectrum has
 a major peak of a photon energy of 1.9 eV and a minor peak
 of an energy of 2.2 eV. Since the electrons and holes for a
 radiative recombination are injected into the levels of 
nanocrystals through avalanche breakdown, the light emission 
voltage can be reduced to 4.0- 4.5 V, which is low enough to be 
applied by a standard Si device.
 
   The authors are grateful to Dr. T. Sakai of the Advanced 
Semiconductor Laboratories and S. Hosoi of the Microelectronics 
Center of Toshiba Corporation for their valuable suggestions and 
technical support in device fabrication. 
They would also like to thank Dr. A. Toriumi, 
M. Ishikawa, and Dr. A. Kurobe of Research and Development 
Center of Toshiba Corporation for their helpful discussion.

\begin{figure}
\caption{ Cross-sectional scheme of the formed device.
}
\label{Fig1}
\end{figure}

\begin{figure}
\caption{Cross sectional HR-TEM lattice image of the a-Si layers that ex
perienced one minute's oxidation.}
\label{Fig2}
\end{figure}

\begin{figure}
\caption{I-V characteristic measured for a Si-NC device that experienced
 one minute of oxidation.}
\label{Fig3}
\end{figure}

\begin{figure}
\caption{EL spectra of a Si-NC device measured at room temperature under
 the reverse bias of 5.0V and the current of 60 mA.}
\label{Fig4}
\end{figure}

\end{document}